\documentstyle[aps,prl,epsfig,multicol,amsmath,amssymb,amsfonts]{revtex}
\draft
\begin{document}

\title{Self-organized and driven phase synchronization in coupled maps} 
\author {Sarika Jalan\footnote{e-mail: sarika@prl.ernet.in} and
R. E. Amritkar\footnote{e-mail: amritkar@prl.ernet.in}}
\address{Physical Research Laboratory, Navrangpura, Ahmedabad 380 009, India.}
\maketitle
  
\begin{abstract}
We study the phase synchronization and cluster formation in coupled
maps on different
networks. We identify two different mechanisms of cluster formation;  
(a) {\it Self-organized} phase synchronization which leads to clusters
with dominant intra-cluster 
couplings and (b) {\it driven} phase synchronization which leads to
clusters with dominant inter-cluster 
couplings. In the novel driven synchronization the nodes of one cluster 
are driven by those of the others. We also discuss the dynamical origin of
these two mechanisms for small networks with two and three nodes.

\end{abstract}


\begin{multicols}{2}

Recently, there is considerable interest in complex systems described by
networks or graphs with complex topology \cite{Strogatz}.
Most networks in the real world consist of dynamical
elements interacting with each other. Thus in order to understand 
properties of such dynamically evolving networks, we study a coupled
map model of different 
networks. Coupled maps show rich phenomenology that arises when
opposing tendencies compete; the nonlinear dynamics of the maps
which in the chaotic regime tends to separate the orbits of different elements,
and the couplings that tend to synchronize them.
Coupled map lattices with nearest neighbor
or short range interactions show interesting spatio-temporal patterns,
and intermittent behavior \cite{CML}. Globally coupled
maps (GCM) where each node is connected with all other nodes, show
interesting synchronized behavior \cite{GCM}.
Ref. \cite{H-Chate} are some of
the papers which shed light on the collective
behavior and synchronization of coupled maps/oscillators with local
and non-local connections on different networks.

In this paper we study the mechanisms for synchronization behaviour of coupled
maps on different networks. In particular, we concentrate on
networks with small number of connections, i.e. the number of connections
($N_c$) are
of the order of the number of nodes ($N$). Our study reveals two different
ways for the formation of synchronized clusters. (a)
Synchronized clusters can be formed because of intra-cluster
couplings. We will refer to this as {\it self-organized
synchronization}. (b)  Synchronized clusters can be formed because of
inter-cluster couplings. Here nodes of one cluster are driven by those
of the others. We will refer to this as {\it driven
synchronization}. We are able to identify ideal clusters of both types,
as well as clusters of the mixed type where both ways of
synchronization contribute
to cluster formation. We will discuss several examples to
illustrate both types of clusters.
Dynamically, our analysis indicates that the self-organized behaviour
has its origin in the
decay term arising due to intra-cluster couplings in the dynamics of
the difference variables while the driven behaviour has its origin in the
cancellation of the inter-cluster
couplings in the dynamics of the difference variables.

Consider a network of $N$ nodes and $N_c$ connections.
Let each node of the
network be assigned a dynamical variable $x^i, i=1,2,\ldots,N$. The
evolution of the dynamical variables is given by
\begin{equation}
x^{i}_{t + 1} = (1 - \epsilon) f( x^{i}_t ) + \frac{\epsilon}{\sum_j C_{i j}}
\sum C_{ij} g( x^{j}_t )
\label{coupleddyn}
\end{equation}
where $x^{i}_n$ is the dynamical variable of the i-th node at
the $t$-th time step, $C$ is the adjacency matrix with elements $C_{ij}$
taking values $1$ or $0$ depending upon whether i and j are
connected or not.  The matrix $C$ is symmetric with diagonal 
elements zero. The function $f(x)$ defines the local 
nonlinear map and 
the function $g(x)$ defines the nature of coupling between the nodes. 
In this paper, we present the results for the local
dynamics given by the logistic map $f(x)=\mu x(1-x)$ and
two types of coupling functions, (i) $g(x) = x$ and (ii) $g(x) = f(x)$.

Synchronization of coupled dynamical systems may be 
defined in various ways \cite{book-syn}.
Perfect synchronization corresponds to
the dynamical variables for different nodes having identical
values. Phase synchronization corresponds to the dynamical variables 
for different nodes having values with some definite relation 
\cite{phase}. For networks with $N_c \sim N$,
we find that perfect synchronization leads to clusters with very
small number of nodes, while phase synchronization gives clusters with large
number of nodes. Here, we concentrate on phase synchronized
clusters. We define the 
phase synchronization as follows \cite{syn}.  Let $n_i$ and $n_j$ denote
the number of times the variables $x^i_t$ and $x^j_t$,
$t=1,2,\ldots,T$ for the nodes $i$ and $j$ show local 
minima during the time interval $T$. Let $n_{ij}$
denote the number of times these local minima match
with each other. We define the phase distance
between the nodes $i$ and $j$ as $d_{ij}=1-2n_{ij}/(n_i+n_j)$. 
Clearly, $d_{ij}=0$
when all minima of variables $x^i$ and $x^j$ match with each other
and $d_{ij}=1$ when none of the minima match.
We say that nodes $i$ and $j$ are phase synchronized if $d_{ij}=0$. 
Also, a cluster of nodes is phase synchronized if all pairs of nodes
of that cluster are phase synchronized.

We find examples of both self-organized and driven types of phase
synchronized clusters in different networks that we have studied.
For small
coupling strengths, we observe turbulent behaviour, i.e. no clusters are
formed, but as the coupling strength increases phase synchronized
clusters are formed. The number and sizes of clusters as well as their
type (self-organized, driven or mixed) depends on the coupling
strength $\epsilon$ as well as the type of coupling function $g(x)$.
For networks with number of connections of the order of $N$, and for
linear coupling $g(x)=x$, we observe self-organized phase synchronized 
clusters for small coupling strengths ($\epsilon\sim 0.18$) and driven
phase synchronized clusters for large coupling strength
with a crossover and reorganization of nodes between the two
types as $\epsilon$ is increased. This behaviour appears
to be approximately independent of the type of network. On the other
hand, for nonlinear coupling $g(x)=f(x)$, we observe a dominant driven 
phase synchronization. In this case, the sizes and number of clusters
depends on the type of network for large $\epsilon$ values.
As noted earlier, in this letter we concentrate only on the
mechanism of cluster formation and other details will be discussed
elsewhere \cite{unpub}.

We now present the numerical results of our model. Starting from random 
initial conditions and after an initial transient, we study the
dynamics of Eq.~(\ref{coupleddyn}), and determine synchronization
behaviour. Fig.~1 shows several examples of clusters illustrating
different behaviors. Fig.~1(a) shows two clusters with an ideal
self-organized phase synchronization. We note that except one
coupling, which must be present since our network is connected, all
other couplings are of intra-cluster type. Fig.~1(b) shows the
opposite behaviour of two clusters with an ideal driven
phase synchronization. Here, all the couplings are of the inter-cluster
type. Fig.~1(c) to 1(e) show mixed behaviour. Fig.~1(c) shows clusters
of different types. The largest two clusters have approximately equal
number of inter-cluster and intra-cluster couplings (mixed type), the next two
clusters have
dominant intra-cluster couplings (self-organized type) while the remaining
three clusters have dominant inter-cluster couplings (driven
type). Also there are several isolated nodes.
Fig.~1(d) shows
clusters where driven behaviour dominates. Fig.~1(e) shows clusters
where self-organized behavior dominates. Fig.~1(f) shows two clusters
of ideal driven type with several isolated nodes. Figs.~1(c) to~1(f)
have isolated nodes which do not belong to
any cluster. These nodes evolve independently, however, some of them
can get attached to some clusters intermittently.

To get a quantitative picture of
the two ways of cluster formation we define two quantities
$f_{inter}$ and $f_{intra}$ as
\begin{mathletters}
\begin{eqnarray}
f_{intra} &=& \frac{N_{intra}}
{N_c}\\
f_{inter} &=& \frac{N_{inter}}
{N_c}
\end{eqnarray}
\end{mathletters}
where $N_{intra}$ is the number of intra-cluster couplings and
$N_{inter}$ is the number of inter-cluster couplings. Couplings between
isolated nodes are not counted in $N_{inter}$.

Figs.~2(a) and 2(b) show both $f_{intra}$ and $f_{inter}$ as a function of
$\epsilon$ for $g(x)=x$ and $g(x)=f(x)$ respectively for the
scale-free networks.
>From Fig.~2(a) ($g(x)=x$), we see that after an initial turbulent phase we get
clusters with large values of $f_{intra}$, i.e. self-organized clusters, for
$\epsilon\underset{\sim}{>} 0.12$.  $f_{intra}$ becomes almost one for $\epsilon\sim 0.18$ 
and then starts decreasing. As $\epsilon$ increases further
$f_{inter}$ starts increasing and there is a crossover and
reorganization of nodes to driven clusters, so that for very large $\epsilon$,
$f_{inter}$ is close to one.  The ideal driven cluster shows
two driven clusters which are anti-phase synchronized with each
other. On the other hand, for $g(x)=f(x)$, Fig.~2(b)
shows that $f_{inter}$, i.e. driven behaviour,
dominates. Fig.~2(c) and~2(d) show similar graphs for network with one 
dimensional nearest neighbor couplings.
The behaviour is similar to that of scale-free
networks except that for $g(x)=f(x)$ and for large $\epsilon$ there is 
almost no synchronization or cluster formation (Fig.~2(d)).

It is interesting to note that the two different ways of cluster
formation are observed even when the variables in the clusters are
evolving chaotically. For $\mu=4$, we find that when three or more
clusters are formed
the largest Lyapunov exponent is positive. When two
clusters (with or without some isolated nodes) are formed largest
Lyapunov exponent can be both positive or negative depending on the
parameter values and the type of coupling. If the
largest Lyapunov exponent is negative, the variables show periodic
behavior with even period \cite{unpub}. For $mu<4$, we find different periods
including odd ones and also two, three or more stable clusters
depending upon the parameters and initial conditions.

Geometrically, the organization of the network into couplings of both
self-organized and driven types is easy to understand for the networks 
with tree structure. A tree can 
always be broken into two or more disjoint clusters with only
intra-cluster couplings by breaking one
or more connections. Clearly, this splitting is not unique. A tree can 
also be divided into two clusters by putting connected nodes into different
clusters. This division is unique and leads to two clusters with only
inter-cluster couplings. For other types of networks splitting into
clusters with ideal intra-cluster or inter-cluster  couplings may not be always
possible. However, clusters with dominant couplings of either
intra-cluster or inter-cluster type is still possible for $N_c\sim N$.
For larger values of $N_c$, typically of the order of $N^2$, a
clear identification of only one type of behavior becomes difficult
and the clusters are mostly of the mixed type. Note that geometrically it is
always possible to get one big cluster spanning almost all the nodes of the 
self-organized type.

A comment on the choice of $T$ used to determine the phase
synchronization. Clearly, $T$ should be large enough to include
several maxima and minima.
On the other hand it should be small enough to include the behavior of
some isolated nodes that get attached to some clusters intermittently
with a time scale of $\tau_s$. We find
that $\tau_s$ is about few thousand iterates. Hence, we chose $T=100$.

We have also studied networks with large $N$ (largest $N$ was
10,000). We can clearly 
identify both self-organized and driven behavior in such
large networks also.

To understand the dynamical origin of the self-organized and driven
phase synchronization let us first consider a network of two
variables, $x^1$ and $x^2$, coupled to each other. 
Synchronization between these two variables will be decided by the
difference variable $x^{s-}=x^1-x^2$. From Eq.~(\ref{coupleddyn}) the
dynamics of $x^{s-}$ is given by
\begin{equation}
x^{s-}_{t+1} = (1-\epsilon) (f(x^1_t)-f(x^2_t)) - \frac{\epsilon}
{2}(g(x^1_t)-g(x^2_t).
\label{diff-self}
\end{equation}
Ref.~\cite{book-syn} discusses synchronization properties of this
network of two variables for $g(x)=f(x)$.
Next consider a network of three variables with both $x^1$ and $x^2$
coupled to $x^3$ and no coupling between $x^1$ and $x^2$. Now the
dynamics of the difference $x^{d-}=x^1-x^2$ is given by 
\begin{equation}
x^{d-}_{t+1} = (1-\epsilon) (f(x^1_t)-f(x^2_t)).
\label{diff-driv}
\end{equation}
It can be shown that there is a critical value of $\epsilon$ above which
the variables $x^1$ and $x^2$ will synchronize, i.e. $x^{d-}$ will
tend to zero. The detailed
dynamics of the above two simple networks and their synchronization
behavior will
be discussed elsewhere \cite{unpub}.

Comparison of Eqs.~(\ref{diff-self}) and~(\ref{diff-driv}) clearly
shows the different dynamical origins of the self-organized and driven
mechanisms. The intra-cluster coupling term which is responsible for
the self-organized behaviour, adds a decay term to the
dynamics of $x^{s-}$ (Eq.~(\ref{diff-self})). On the other hand, the
inter-cluster coupling terms, which are responsible for the driven behaviour,
cancel out and do not add any term to the
dynamics of $x^{d-}$ (Eq.~(\ref{diff-driv})).  We feel that for larger networks
also similar mechanisms as in Eqs.~(\ref{diff-self})
and~(\ref{diff-driv}) are responsible for the cluster formation of the
self-organized and driven type.

There are several examples of self-organized and driven behavior in
naturally occurring systems. Self-organized behavior is common
and is easily observed. Examples are social, ethnic and religious
groups, political groups, cartel of
industries and countries, herds of animals and flocks of birds,
different dynamical transitions such as self-organized criticality etc. The
driven behavior is probably not so common \cite{note1}. An interesting
example is the
behavior of fans during a match between traditional rivals. 
Before the match the fans may act as individuals
(turbulent behavior). During the match, when the
match reaches a feverish pitch, i.e. the strength of the interaction
increases, the fans are likely to form two driven phase synchronized groups.
The response of each group depends on that of the other and
is normally
anti-phase synchronized with the other. Another example is the formation of
opposite ethnic groups as in Bosnia. 

In this letter we have presented results for the case where the local
dynamics is governed by logistic map and couplings $g(x)=x$ and
$f(x)$ for the scale free networks and one-d lattice with nearest neighbor
couplings. We have also studied several
other maps (circle map, tent map etc.), other types of couplings
and different networks (two-d lattice with nearest neighbor
couplings, small world networks,
Caley tree, random networks,
bipartite networks). 
We find similar behaviour and
are able to identify self-organized and driven behaviour. 

To conclude we have investigated the mechanism of cluster formation
in coupled maps on different networks. We are able to identify two
distinct ways of cluster formation, namely
self-organized and driven phase
synchronization. Self-organized synchronization is
characterized by dominant intra-cluster couplings while driven behavior
is characterized by inter-cluster couplings.
The two ways of cluster formation are clearly seen for networks with
small number of
couplings ($N_c\sim N$) but are difficult to identify as the
number of couplings increases and becomes of the order of $N^2$.
Dynamically, the examples of small networks show that the
self-organized behaviour occurs because of
the intra-cluster couplings introducing a decay term in the difference
variables while the driven behaviour occurs because of the inter-cluster
couplings cancelling out.  

One of us (REA) thanks Professor H. Kanz for useful
discussions and Max-Plank Institute for
the Physics of Complex Systems, Dresden for hospitality.

\end{multicols}

\newpage

\begin{figure}
\begin{center}
\epsfxsize 15cm \epsfysize 10cm
\epsfbox[62 362 562 715]{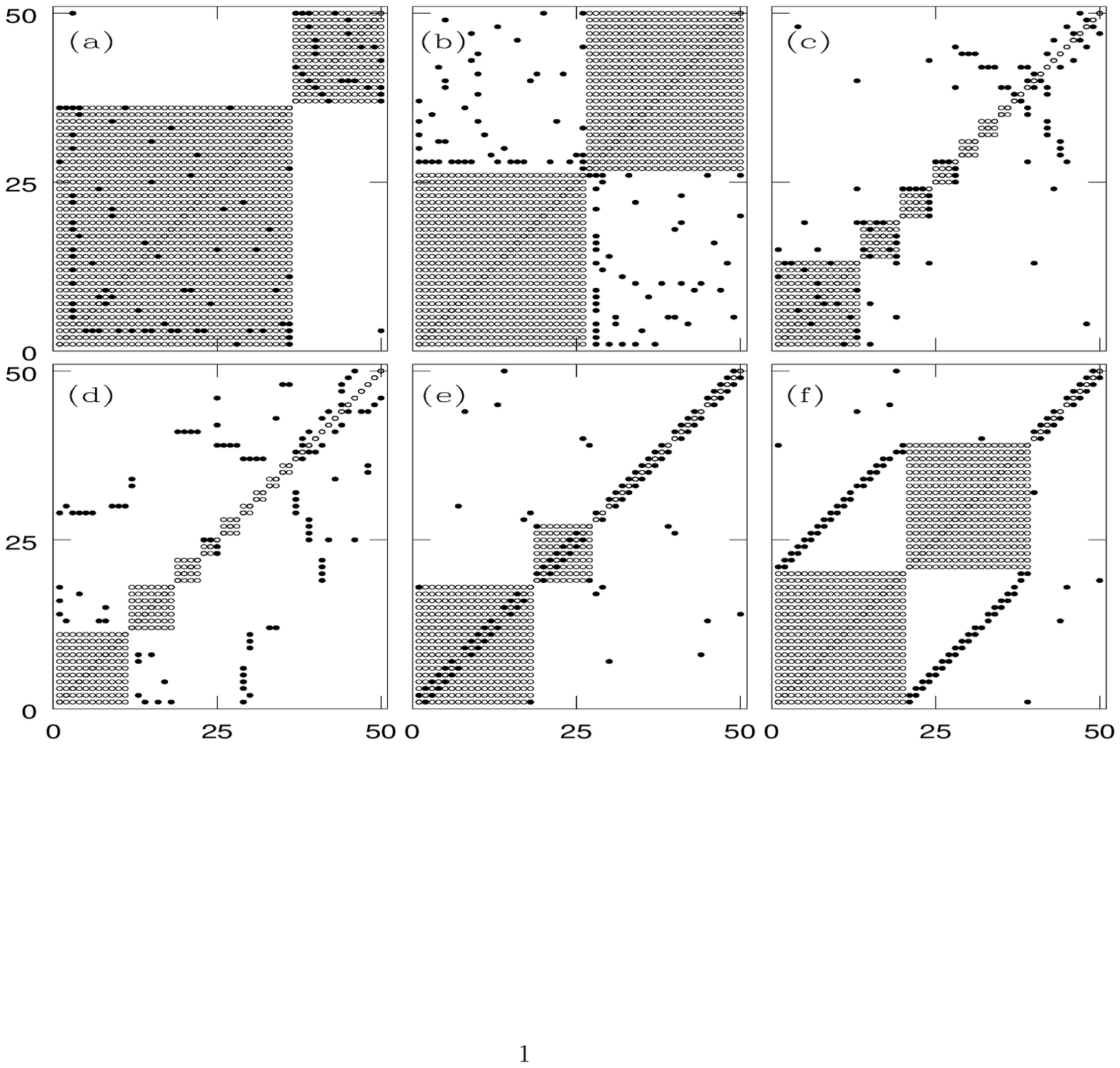}
\caption {The figure shows several examples illustrating the self-organized and
driven phase synchronization. The examples are chosen to demonstrate
the two different ways of obtaining synchronized clusters and the
variety of clusters that
are formed. All the figures show node verses node diagram for
$N=N_C=50$. After an initial transient (about 2000 iterates) phase
synchronized clusters are
studied for $T=100$. The logistic map parameter $\mu=4$. The solid
circles show that the two corresponding nodes are coupled and the open
circles show that the corresponding nodes are phase
synchronized. In each case the node numbers are reorganized so that
nodes belonging to the same cluster are
numbered consecutively. (a) Figure show an ideal self-organized phase
synchronization for scale
free network for $g(x)=x$ and$\epsilon=0.15$. (b) An ideal driven phase
synchronization for scale free network for $g(x)=x$ and
$\epsilon=0.85$. (c) Mixed behavior for scale free network for
$g(x)=f(x)$ and $\epsilon=0.61$. (d) A dominant driven behavior for scale
free network for $g(x)=f(x)$ and $\epsilon=0.87$. (e) A dominant
self-organized behavior for 1-d lattice with nearest neighbor couplings for
$g(x)=x$ and $\epsilon=0.14$. (f) An ideal driven behavior with several
isolated nodes for 1-d lattice with nearest neighbor couplings for $g(x)=f(x)$
and $\epsilon=0.15$. The
scale free networks were generated starting with $N_0=1$ nodes 
and adding one node with $m=1$ couplings at each stage of
the growth of the lattice with probability of connecting
to a node being proportional to the degree of the node (see
Ref. [9] for details).}
\end{center}
\end{figure}

\begin{figure}
\begin{center}
\epsfxsize 15cm \epsfysize 10cm
\epsfbox[70 350 564 723]{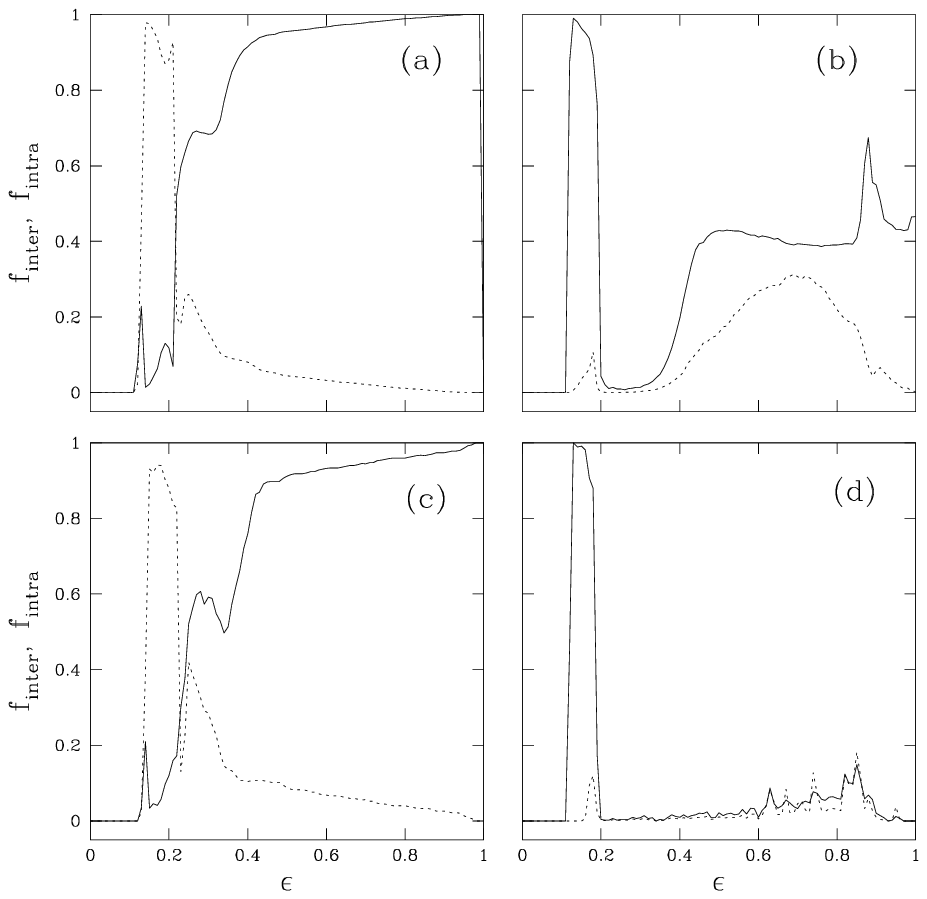}
\caption{The fraction of intra-cluster and inter-cluster couplings,
$f_{inter}$ (solid line) and $f_{intra}$ (dashed line) are shown
as a function of the coupling
strength $\epsilon$. Figures (a) and (b) are for the scale free network 
for $g(x)=x$ and $g(x)=f(x)$ respectively. Figures (c) and (d) are
for the 1-d network with nearest neighbor couplings for $g(x)=x$ and $g(x)=f(x)$
respectively. The figures are obtained by averaging over 20
realizations of a network and 50 random initial conditions for each realization.
}
\end{center}
\end{figure}


\begin{thebibliography}{99}
\bibitem{note1} The two sub-lattice antiferro-magnetism in statistical
mechanics is also of the driven type, but it is an equilibrium
phenomenon, while the driven clusters that we have are of dynamical origin.
\end{thebibliography}
\end{document}